\documentstyle[multicol,prl,aps,epsf,psfig,epsfig]{revtex}

\begin{document}
\draft

\title{A Cooperative Stochastic Model of Gene Expression}

\author{Siddhartha Roy$^1$, Indrani Bose$^2$ and
Subhrangshu Sekhar Manna$^3$}

\address{$^1$ Department of Biophysics, Bose Institute, 
P1/12, C.I.T. Scheme VII M, Calcutta-700054, India \\
$^2$ Department of Physics, Bose Institute, 93/1,
A.P.C. Road, Calcutta- 700009, India \\
$^3$ Satyendra Nath Bose National Centre for Basic Sciences, 
Block-JD, Sector-III, Salt Lake, Calcutta 700098,
India
}

\maketitle

\begin{abstract}
Recent experiments at the level of a single cell have
shown that gene expression
occurs in abrupt stochastic bursts. Further, in an
ensemble of cells, the levels
of proteins produced have a bimodal distribution. In a
large fraction of cells,
the gene expression is either off or has a high value.
We propose a stochastic
model of gene expression the essential features of
which are stochasticity
and cooperative binding of RNA polymerase. The model
can reproduce the bimodal
behaviour seen in experiments.
\end{abstract}

\pacs {Key Words: Gene expression. Stochastic
transcription, Bimodal distribution}

\begin{multicols}{2}\narrowtext

Gene expression is a fundamental and important
biological process in a cell.
Genes are part of DNA molecules and determine the
structure of functional molecules such as RNAs and
proteins. In each cell, at any instant of time, only a
subset of genes present is active in directing
RNA/protein synthesis. The gene expression is 'on' in
such a case. The information present in the gene is
expressed in the following manner. In the first step
of gene expression, the sequence along one of the
strands of the DNA molecule is copied or transcribed
in a RNA molecule (mRNA). The sequence of mRNA
molecules is then translated into the sequence of
amino acids, which in turn determines the functional
nature of the protein molecule produced. The rate and
temporal sequence of gene expression is responsible
for many aspects of biology. In the large majority of
cases, the regulation of gene expression occurs at the
level of transcription and hence an in-depth
understanding of transcription regulation is a central
focus of biology \cite{1}.

Recent experiments (see Appendix A), provide evidence
that gene expression occurs in abrupt stochastic
bursts at the level of an individual cell\cite{2,3,4}.
Also, in many cases, in a population of cells the
levels of proteins produced are 
distributed in a bimodal manner implying that in a
large fraction of cells 
the gene expression is either off or has a high
value\cite{5}. In this paper, we propose a stochastic
model of gene expression which provides 
a possible explanation of the observed bimodal
behaviour. 

Genes are transcribed into mRNA by an enzyme called
RNA polymerase (RNAP). The
process is initiated with the binding of RNAP to a
site called promoter, usually
near the beginning of the transcribed sequence. After
the initial binding and
subsequent conformational changes, the enzyme begins
synthesis of the RNA chain
and gradually translates along the DNA. The initial
binding of RNAP to a promoter
can be prevented by the binding of a regulatory
protein (R) to an overlapping
segment of DNA (called operator) resulting in a
turning off of mRNA production.
There is a finite probability that the bound R
molecule dissociates from the
operator at any instant of time. RNAP molecule then
has a certain probability
of binding to the promoter and initiating
transcription.

\begin{figure}
  \centerline{\epsfig{file=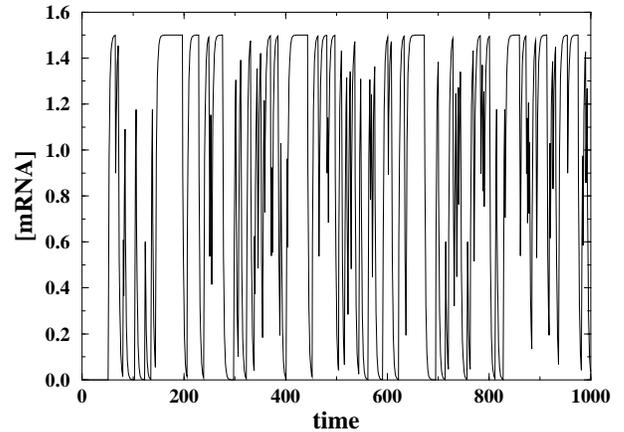, width=8cm}}
  \vspace*{0.5cm}
\caption{
Concentration of mRNA molecules {[}mRNA{]} in
arbitrary units as a function
of time t. The parameter values are $p_1$ = 0.5, $p_2$
= 0.5,
$p_3$ = 0.3, $p_4$ = 0.85, $p_5$ =0.05 and $\mu$=0.4.
}
\end{figure}

Each of the possibilities described above actually
involves a series of physico-chemical
processes, a detailed characterization of which is not
required for the model
of gene expression that we propose here. We represent
a gene by a one-dimensional
lattice of n+2 sites. The first two sites represent
the operator and promoter
respectively. The lattice is a coarse-grained
description of an actual gene.
In reality the operator and promoter regions may
extend over a certain number
of base pairs in the DNA and they can be overlapping
or not. In our model they
are represented as single sites. Each of the other
sites in the lattice represents
a finite number of base-pairs in the DNA molecule.

The different physico-chemical processes are lumped together into a few simple
events which are random in nature. This lumping together avoids unnecessary
complexity that has no bearing on the basic nature of the process. The operator
(O) and promoter (P) together can be in four possible configurations : 10, 01,
00 and 11. The numbers `1' and `0' stand for `occupied' and `unoccupied'. The
configuration $ij$ describes the occupation status of O ($i$) and P ($j$). For 
example, the configuration 10 corresponds to O being occupied by a R molecule 
and P being unoccupied. Similarly, in the configuration 01, O is unoccupied and 
P is occupied by a RNAP molecule. Binding of R and RNAP molecules are mutually 
exclusive so that the configuration 11 is strictly prohibited. Given a 00 
configuration at time $t$, the transition probabilities to configurations 10 and 
01 at time $t+1$ are $p_1$ and $p_2$ respectively. The probability of remaining
in the configuration 00 is $1-p_1-p_2$ . A 10 configuration at time $t$ goes to 
a 00 configuration at time $t+1$ with probability $p_3$ and remains unchanged 
with probability $1-p_3$ .We have assumed all the probabilities to be 
time-independent. The justification for this approximation is that the number of 
free R and RNAP molecules in the cell are typically one or two orders of magnitude 
higher than the number of DNA sites they occupy. The RNAP molecule once bound to 
the promoter initiates transcription in the next time step, i.e., the 01 
configuration makes a transition to a 00 configuration with probability 1. The 
motion of RNAP is in the forward direction and the molecule covers a unit distance 
(the distance between two successive lattice sites) in each time step. Once the 
molecule reaches the last site of the lattice the transcription ends and a mRNA 
is synthesized.

 The second major feature of our model is the cooperative binding of RNAP to
the promoter, when an adjacent RNAP molecule is present. This implies that 
there is a higher probability of binding of RNAP to the promoter in one time 
step if another RNAP molecule is present at the site next to the promoter.In our 
model, the probability of cooperative binding of RNAP is $p_4$ which is larger 
than $p_2$. The probabilities $p_1$ and $1-p_1-p_2$ are changed to new values 
$p_5$ and $1-p_4-p_5$ respectively. Degradation of mRNA is taken into account 
by assuming the decay rate to be given by $\mu$N where N is the number of mRNAs 
present at time $t$. The number of mRNAs produced as a function of time is studied 
by Monte Carlo simulation.For the sake of simplicity, we have not tried to simulate 
protein levels or enzymatic products thereof, i.e., we study gene expression upto
the level of transcription (mRNA synthesis). Since the number of protein molecules 
and converted products should be proportional to the mRNA present, no loss of 
generality is introduced by this simplification. The lattice consists of 52 sites 
($n$=50). Stochastic events are simulated with the help of a random number generator.
The updating rule of our cellular automaton (CA) model is that in each time step 
$t$ the occupation status (0 or 1) of each site (except for the O site) at time 
$t - 1$ is transferred to the nearest-neighbour site towards the right. If
the $(n+2)$-th , i.e. , the last site is 1 at $t - 1$, a mRNA is synthesized at $t$
and the number of mRNAs increases by 1. In the same time step, the configuration 
$ij$ of OP is determined with the probabilities already specified. Thus in each
time step, the RNAP molecule, if present on the gene, moves forward by unit lattice
distance (progression of transcription) followed by the updating of the OP
configuration. Figure 1 shows the concentration {[}mRNA{]} of mRNA molecules
in the cell as a function of time for the parameter values $p_1$ = 0.5, $p_2$ =
0.5, $p_3$ = 0.3, $p_4$ = 0.85, $p_5$ = 0.05 and $\mu$ = 0.4. Note that
an almost four-fold increase in the probability of RNAP binding is assumed due
to cooperativity. The stochastic nature of the gene expression is evident from
the figure with random intervals between the bursts of activity. One also notices
the presence of several bursts of large size. It is important to emphasize that
the frequency of transitions between high and low expression levels is a function
of the parameter values chosen and may be low for certain parameter values.
For the probability values considered, the two predominantly favourable states 
are when the gene expression is off (state 1) and when a large amount of gene
expression takes place (state 2). In the absence of RNAP, state 1 has greater
weightage but with the chance binding of RNAP to the promoter (probability $p_2$ for
this is small), the weight shifts to state 2 until another stochastic
event terminates cooperative binding and the gene reverts to state 1. The
probability of obtaining a train of $N$ successive transcribing RNAP molecules is
$p_2p^{N-1}_4(1-p_4)$. This is the geometric distribution function and the
mean and the variance of the distribution are given by $p_2/(1-p_4)$ and
$p_2(1+p_4-p_2)/(1- p_4)^2$ respectively.

For the probability values already specified, the simulation has been repeated
for an ensemble of 3000 cells. For each cell, the time evolution is upto 10,000
time steps. Figure 2 shows the distribution of the number $N(m)$ of cells versus
the fraction $m$ of the maximal number of mRNA molecules produced after 10,000
time steps. Two distinct peaks are seen corresponding to zero and maximal gene
expression respectively. Such a bimodal distribution occurs over a wide range
of parameter values. Figure 3 shows the distribution for parameter values 
$p_1$=0.7, $p_2$=0.2, $p_3$ = 0.7, $p_4$ = 0.85, $p_5$ =0.05 and $\mu$=0.5.
For the same set of parameter values but with $p_3$ = 0.1, the bimodal
distribution is lost and one gets a single prominent peak corresponding to maximal
gene expression. Distributions with several peaks of random heights are obtained
when the parameter values do not produce the effect. The full parameter
space describing the three different regions of unimodal, bimodal and multi-peak
distributions has not been explored in detail as yet. The transition from one
region to another is in a broad sense like a phase transition.Since the distribution 
of transcribing RNAPs is bimodal in nature, many results like the distribution 
of time intervals in between bursts of gene expression can be written down from the
stochastic theory of such distributions \cite{6}.

In summary, we have proposed a stochastic model which
can reproduce the bimodal distribution in gene
expression observed in recent experiments. We have
suggested that the stochastic nature of transcription
coupled with RNAP binding cooperativity may result in
discontinuous levels of gene expression and consequent
bimodal distribution of expressed protein levels, as
observed in a number of experiments. To our knowledge,
no stochastic mechanism of bimodal distribution has
been offered so far. Increasing emphasis on the
stochastic nature of the developmental switches
operating at the level of transcription suggests that
the bimodal distribution of protein levels may have a
role to play in such mechanisms \cite{7}.

\begin{figure}
  \centerline{\epsfig{file=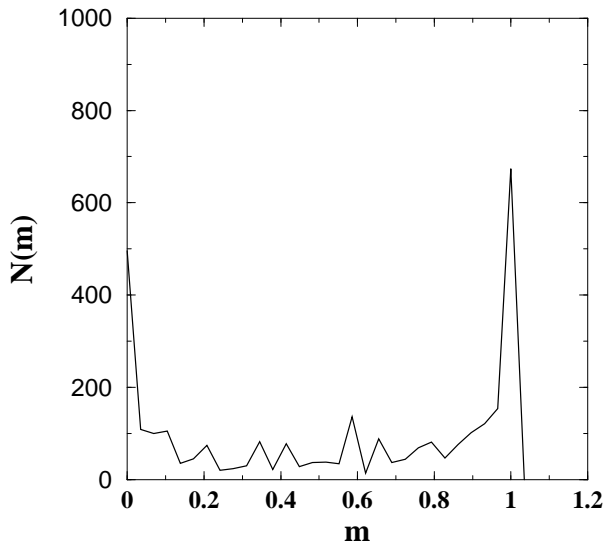, width=8cm}}
  \vspace*{0.5cm}
\caption{
Distribution of no. N(m) of cells expressing fraction
m of maximal number of
mRNA after 10,000 time steps. The total number of
cells is 3000. The
parameter values are $p_1=p_2$=0.5, $p_3$=0.3,
$p_4$=0.85, $p_5$=0.05 and $\mu$=0.4.
}
\end{figure}

\vskip 0.4 cm
{\centerline {\bf {\underline {Appendix A}}}
\vskip 0.4 cm

In this Appendix, we discuss the various biological
aspects of the problem studied in this paper.

Biological variability is a product of interaction of
genes with the environment. With the advent of rapid
genome sequencing methods and remarkable success 
in sequencing genomes from many organisms, the thrust is
now gradually shifting to the functional aspects of
information present in the genome. The genome of an
organism is a storehouse of sequential information
contained in all the genes specific to that organism.
Through gene expression, the sequential information
determines the structure of functional molecules like
RNAs and proteins.Since the advent of molecular
biology, the regulation of gene expression has
been studied in solution or in an ensemble of cells
where an average property
is measured. This mode of study was necessary, as it
was difficult to obtain
information at the level of an individual cell. A
complete understanding of
cellular processes, however, needs an appreciation of
events at the level of
an individual cell and extrapolation to an ensemble of
cells. 
\begin{figure}
  \centerline{\epsfig{file=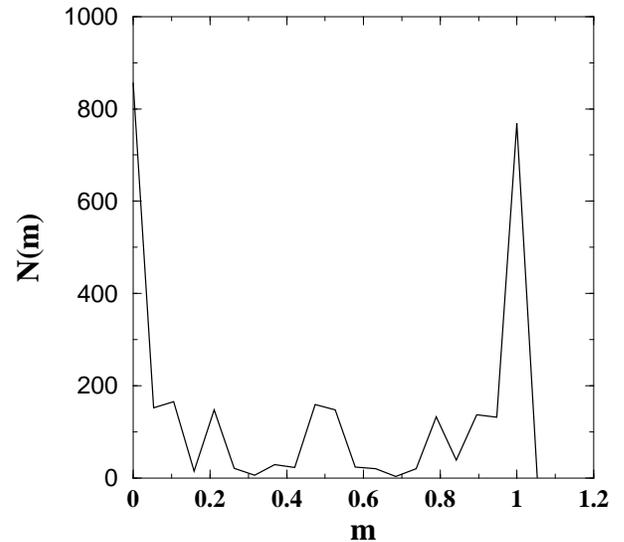, width=8cm}}
  \vspace*{0.5cm}
\caption{
Distribution of no. N(m) of cells expressing fraction
m of maximal number of mRNA
after 10,000 time steps. The total number of cells is
3000. The parameter
values are $p_1$=0.7, $p_2$=0.2, $p_3$=0.7,
$p_4$=0.85, $p_5$=0.05 and $\mu$=0.5.
}
\end{figure}
Recent advances
have made it possible to study processes within a
single cell unmasked by ensemble
averaging \cite{8}. The simplest event one can study
at the individual cell
level is that of the expression of a reporter gene
such as lacZ and GFP. In
the former case, the end product is an enzyme $\beta$
- galactosidase
which is capable of hydrolyzing a non-coloured
substrate to a coloured product.
In the latter case the protein itself is fluorescent.
Hence, the gene expression
can be directly studied either colorimetrically or
flurometrically at the level
of an individual cell. Recent experiments using such
techniques, provide evidence
that gene expression occurs in abrupt stochastic
bursts at the level of an individual
cell \cite{2,3,4}. The stochastic nature of gene
expression is also evident
when levels of $\beta$-galactosidase were examined in
an ensemble of
cells. Levels of $\beta$-galactosidase are distributed
in a bimodal manner,
in a large fraction of cells the gene expression is
either off or has a high
value \cite{5}.

Some theories have been proposed so far to explain the
so-called `all or none'
phenomenon in gene expression. These theories are
mostly based on an auto-catalytic
feedback mechanism, synthesis of the gene product
gives rise to the transport
or production of an activator molecule \cite{5,8,9}.
While such processes are 
certainly possible, the bimodal distribution is a much
more general phenomenon
and has now been found in many types of cells, from
bacterial to eukaryotic
and for different types of promoters \cite{2,3,4}.

The two major features of the model of gene expression
that we have proposed in this paper are stochasticity
and cooperative binding of RNA polymerase. As already
explained in the paper, the different physico-chemical
processes associated with gene expression are lumped
together into a few simple events which are random in
nature. To give an example, for many prokaryotic
promoters, there is a two-step reaction scheme in
which a RNAP open complex is formed preceded by the
formation of a closed complex. RNAP initiates
transcription only from the open complex. The
isomerization step is rate-limiting in many cases
\cite{10}. We define the on-rate of RNAP as the
composite of several steps with the final attainment
of the open complex. The cooperative binding of RNAP
to the promoter in our model implies that there is a
higher probability of binding of RNAP to the promoter
in one time step if another RNAP molecule is present
at the site next to the promoter. Although such
binding cooperativity has not been studied in
prokaryotic polymerases, it has been demonstrated in
polio-virus RNA-dependent RNA polymerase
  \cite{11}.
Cooperative binding of proteins to DNA is now well
established. In most cases
of regulatory proteins, the binding cooperativity is
mediated through protein-protein
interaction although increasing evidence of DNA
mediated effects are being reported
\cite{12}. In the case of RNAP binding to promoters,
however, there are now
widespread reports of transcription generated increase
in negative supercoiling
with consequent increase in rate of transcription
\cite{13,14}. In many promoters,
the transcription initiation is sensitive to the
supercoiling status of the
DNA. It has been reported that transcription generates
increased negative supercoiling
through several hundred base pairs \cite{15}. Thus it
is entirely plausible
and likely that active transcription downstream of the
promoter site may lead
to increased binding of RNAP and open-complex
formation. One can also envisage
other mechanisms for generating this kind of
cooperativity. For example, if
the polymerase-generated negative supercoiling (after
initial movement) inhibits
binding of the repressor, it would effectively
increase polymerase binding probability.

Transcription is one of the most important events in
the life-cycle of a cell. The temporal sequence of
events occurring during transcription is of
utmost importance in its understanding and has been
studied extensively. The general description of
transcription as well as other cellular events have
tended to be deterministic in nature. In the cell
there are only a few DNA molecules and a few molecules
of free RNAP. It is likely that the number of
molecules in the cell is not high enough so that a
deterministic description of this small ensemble is
correct. At the level of a single cell, probabilistic
descriptions are more appropriate. Increasingly,
probabilistic descriptions of cellular events,
including transcription are being offered
\cite{16,17,18}.

Electronic addressies: sidroy@vsnl.com, \\
indrani@boseinst.ernet.in,
manna@boson.bose.res.in

}
\end{multicols}


\begin{thebibliography}{10}
\bibitem{1}Genes V by B. Levin (Oxford University
Press, New York 1994)
\bibitem{2}G. Zlokarnik, P.A. Negulescu, T.E. Knapp,
L. Mere, N. Burres, L. Feng, M. Whitney,
K. Roemer and R.Y. Tsien, Science 279, 84 (1998)
\bibitem{3}P. A. Negulescu, N. Shastri and M.D.
Cahalan, Proc. Natl. Acad. Sci. 91, 2873
(1994)
\bibitem{4}J. Karttunen and N. Shastri, Proc. Natl.
Acad. Sci. 88, 3972 (1991)
\bibitem{5}A. Novick and M. Weiner, Proc. Natl. Acad.
Sci. 43, 553 (1957)
\bibitem{6}An Introduction to Probability Theory and
its Applications by W.Feller (Wiley
Eastern Limited 1984)
\bibitem{7}D.A. Hume, Blood 96, 2323 (2000)
\bibitem{8}M.T.Beckman and K. Kirkegaard, J. Biol.
Chem. 273, 6724 (1998)
\bibitem{9}T.A. Carrier and J.D. Keasling, J. Theor.
Biol. 201, 25 (1999)
\bibitem{10}W.R. McClure, C.L. Cech and D.E. Johnston,
J.Biol. Chem. 253, 8941 (1978)
\bibitem{11}D. A. Siegele and J.C. Hu, Proc. Natl.
Acad. Sci. 94, 8168 (1997)
\bibitem{12}S. Adhya, Ann. Rev. Genetics 23, 227
(1989)
\bibitem{13}K.Y. Rhee, M. Opel, E. Ito, S. Hung, S.M.
Arfin and G.W. Hatfield, Proc. Natl.
Acad. Sci. 96, 14294 (1999)
\bibitem{14}S.D. Sheridan, C.J. Benham and G. W.
Hatfield, J. Biol. Chem. 273, 21298 (1998)
\bibitem{15}A.S. Krasilnikov, A. Podtelezhnikov, A.
Vologodskii and S.M. Mirkin, J. Mol.
Biol. 292, 1149 (1999)
\bibitem{16}H.H.Mcadams and A. Arkin, Proc. Natl.
Acad. Sci. 94, 814 (1997)
\bibitem{17}H.H. McAdams and A. Arkin, Trends in
Genetics 15, 65 (1999)
\bibitem{18}A.M.Kierzek,J.Zaim and P.Zielenkiewicz,
preprint
\end{thebibliography}
\end{document}